\documentclass[twocolumn]{aastex61}
\usepackage{verbatim}
\usepackage{gensymb}
\usepackage{graphicx}
\usepackage{array}
\usepackage{booktabs}
\usepackage{threeparttable}
\usepackage{natbib}
\usepackage{bm}
\usepackage{amsmath}
\newcolumntype{+}{>{\global\let\currentrowstyle\relax}}
\newcolumntype{^}{>{\currentrowstyle}}

\begin{document}

\title{Detecting Weak Spectral Lines in Interferometric Data through Matched Filtering}

\correspondingauthor{Ryan A. Loomis}
\email{rloomis@cfa.harvard.edu}

\author{Ryan A. Loomis}
\affil{Harvard-Smithsonian Center for Astrophysics, Cambridge, MA 02138}

\author{Karin I. {\"O}berg}
\affil{Harvard-Smithsonian Center for Astrophysics, Cambridge, MA 02138}

\author{Sean M. Andrews}
\affiliation{Harvard-Smithsonian Center for Astrophysics, Cambridge, MA 02138}

\author{Catherine Walsh}
\affiliation{Leiden Observatory, Leiden University, P.O. Box 9531, 2300 RA Leiden, The Netherlands}
\affiliation{School of Physics and Astronomy, University of Leeds, Leeds LS2 9JT, UK}

\author{Ian Czekala}
\altaffiliation{KIPAC Postdoctoral Fellow}
\affiliation{Harvard-Smithsonian Center for Astrophysics, Cambridge, MA 02138}
\affiliation{Stanford University, 452 Lomita Mall, Stanford, CA, 94305}

\author{Jane Huang}
\affiliation{Harvard-Smithsonian Center for Astrophysics, Cambridge, MA 02138}

\author{Katherine A. Rosenfeld}
\affiliation{Harvard-Smithsonian Center for Astrophysics, Cambridge, MA 02138}

\begin{abstract}
Modern radio interferometers enable observations of spectral lines with unprecedented spatial resolution and sensitivity. In spite of these technical advances, many lines of interest are still at best weakly detected and therefore necessitate detection and analysis techniques specialized for the low signal-to-noise ratio (SNR) regime. Matched filters can leverage knowledge of the source structure and kinematics to increase sensitivity of spectral line observations. Application of the filter in the native Fourier domain improves SNR while simultaneously avoiding the computational cost and ambiguities associated with imaging, making matched filtering a fast and robust method for weak spectral line detection.  We demonstrate how an approximate matched filter can be constructed from a previously observed line or from a model of the source, and we show how this filter can be used to robustly infer a detection significance for weak spectral lines. When applied to ALMA Cycle 2 observations of CH$_{3}$OH in the protoplanetary disk around TW Hya, the technique yields a $\approx$53$\%$ SNR boost over aperture-based spectral extraction methods, and we show that an even higher boost will be achieved for observations at higher spatial resolution. A Python-based open-source implementation of this technique is available under the MIT license at \url{https://github.com/AstroChem/VISIBLE}.
\end{abstract}

    %As previously mentioned, the $\chi^{2}$ interpretation of a matched filter applied at the systemic velocity suggests that we could have simply computed a $\chi^{2}$ instead of computing the full convolution. The advantage of a full spectrum application of the matched filter, however, is that we are able to observe the entire noise-suppressed spectrum. In the regime of current interferometers, where a dataset contains hundreds of thousands of visibilities but each visibility is severely noise dominated, reduced $\chi^{2}$ approaches 1 no matter what model is used. Thus without computing $\chi^{2}$ for many models it is difficult to determine the significance of a weak line detection. In contrast, the convolution of a matched filter returns a spectrum that naturally visualizes the noise floor, and also allows for possible detection of unknown species when applied to large broadband datasets such as line surveys.

\section{Introduction}
    The rich spatio-kinematic information that radio interferometric datasets can provide for molecular spectral lines is crucial for studying the astrophysical and chemical processes occurring in host sources. The broadband capabilities of modern interferometers allow many spectral lines to be observed in a single correlator setup, enabling astronomers to simultaneously trace multiple astrophysical phenomena, or undertake unbiased line surveys to work toward complete molecular inventories \citep[e.g.][]{Jorgensen_2011, Coutens_2016, Muller_2016}. Within these datasets, many scientifically interesting lines may be weak or not detected due to low column densities or intrinsically low line strengths. Finding these lines and robustly assessing their strength is key to achieving many science goals.

\begin{figure*}[ht!]
    \centering
    \includegraphics[width=0.7\textwidth]{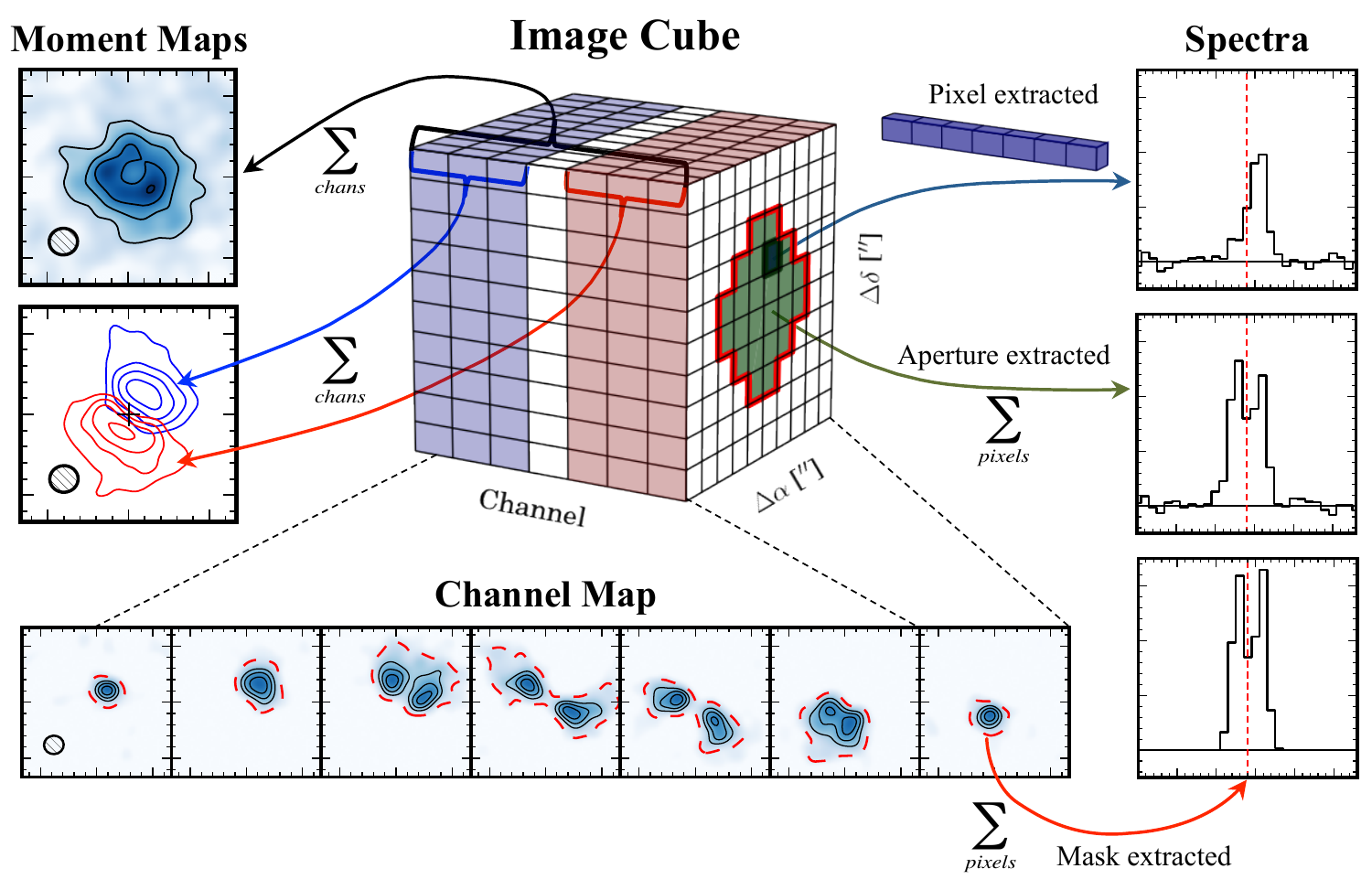}
    \caption{{\small A diagram illustrating the multiple ways of viewing an image cube. Counter-clockwise from the top-left: Velocity-integrated moment maps, made by integrating slices of the cube along the frequency axis; channel maps, where each panel corresponds to a channel of the cube; spectra, generated from top to bottom from a single pixel, integrated over an aperture, and integrated using a matched spatio-kinematic mask (dashed red contours in channel maps). The synthesized beam is shown in the lower left of the moment and channel maps.}  \label{Figure 1}}
    \end{figure*}

    Resolved interferometric observations pose special challenges to detecting weak spectral lines. Radio interferometers measure visibilities, samples of the Fourier transform of the distribution of emission intensities from an astrophysical source at discrete spatial and spectral frequencies. These visibilities are then Fourier inverted and deconvolved with a routine such as \texttt{CLEAN} \citep{Hogbom_1974} to create an image cube. As shown in Fig. \ref{Figure 1}, this image cube consists of a series of images (channel maps) of the emission intensity distribution in distinct spectral frequency bins, which correspond to projected radial velocity bins. In the simplest line detection scenario, emission is directly observed in these channel maps. 

    When emission is too weak to be directly visible in the channel maps, the image cube might be manipulated in a variety of ways to increase SNR. Spectra can be extracted from the cube, and moment maps can be generated by collapsing the cube along the spectral axis, illustrated in Fig. \ref{Figure 1}. All spectral extraction approaches incorporate a spatial mask. If the source is unresolved, a single pixel-extracted spectrum will contain all available information. In cases where the emission is extended and spatially resolved, the simplest mask that contains all emission is an aperture drawn around the source. Such a mask rarely results in spectra with optimal SNR, however. In sources with complex spatio-kinematic patterns, due to e.g. bulk rotation, emission may `move' across the channel maps. The aperture mask is then larger than the emitting area in any given channel, adding noise to the extracted spectrum. To combat this, a spatio-kinematic mask specifically tailored to the structure of the source may be used to reduce the amount of added noise \citep[e.g.][]{Dutrey_2007, Oberg_2015, Loomis_2015, Yen_2016}. 
    
    The application of spatio-kinematic masks to spectral image cubes has already enabled new science, but there are both computational and interpretive challenges when attempting to extend this technique to detect weak lines. First, the observed visibilities must be imaged, a non-trivial computational cost for high resolution observations or spectral surveys with large bandwidths. Second, when the visibilities are Fourier inverted, the PSF is oversampled with pixels to reduce imaging artifacts. This introduces a spatial covariance between pixels on the scale of the beam, making statistical interpretations of extracted spectra difficult. Finally, tailored spatio-kinematic masks reduce added noise but sacrifice a meaningful spectral baseline, making robust weak line detection difficult unless more complicated bootstrapping approaches are taken to establish a false positive rate \citep{Barenfeld_2016}.

    These obstacles can be overcome while retaining the benefits of the spatio-kinematic approach by applying a matched filter directly to the observed visibilities. When the shape of a signal is known, the optimal linear filter for signal extraction is a matched filter, equivalent to the known signal with a normalization constant. Cross-correlating a noisy signal with this filter maximizes the output SNR. This approach is used extensively in digital signal processing; prominent examples include RADAR \cite[e.g.][ and references therein]{Woodward_1953, Cumming_2005}, source detection in imaging surveys \citep[e.g.][]{Bertin_1996, Bertin_2001, Meillier_2016, Herenz_2017, Zackay_2017}, gravitational wave detection \citep[e.g.][]{Owen_Sathyaprakash_1999, Schutz_1999, Abbott_2016}, and exoplanet detection through direct imaging \citep{Ruffio_2017}.
    
    Because matched filters are simply cross-correlated with the data, they can be easily applied in the Fourier domain to provide a fast and unbiased approach to weak line detection over broad bandwidths. An image-plane spectral extraction mask reduces unnecessary noise contributions by incorporating estimated spatio-kinematic (and correspondingly, interferometric phase) information into the extracted spectrum. Similarly, matched filtering quantitatively combines both amplitude and phase information of the observed visibilities into a robust detection probability. Line detection directly in the visibilities both avoids the high computational expense of fully imaging wide-bandwidth datasets and retains a straightforward statistical interpretation of detection significance.
    
    In this paper, we describe how to construct and apply a matched filter to interferometric spectral line data and demonstrate the method on observations from the Atacama Large Millimeter/sub-millimeter Array (ALMA). In \S2, we provide an overview of matched filtering and detail the steps of the method. In \S3, we apply the technique to ALMA Cycle 2 observations from \cite{Walsh_2016} of CH$_{3}$OH in a protoplanetary disk. In \S4, we discuss how much SNR boost might be expected for a given dataset, compare the technique to other methods, and suggest applications where matched filtering may prove useful. A summary is given in \S5. Formulas to approximate the expected SNR boost are derived in Appendices \ref{appendix:A} and \ref{appendix:B}, derivations of noise covariance matrices for correlated channels are presented in Appendix \ref{appendix:C}, and details of the example filters used in the paper are given in Appendix \ref{appendix:D}.

\section{Method}
    In this section we first present a brief overview of the principles behind matched filtering, introducing the one-dimensional matched filter. The one-dimensional approach is then easily extended to higher dimensional problems such as searching for signals within an image or image cube \citep[e.g.][]{Bertin_1996, Feng_2017, Herenz_2017, White_2017}. We present here a novel method to apply matched filters in the native measurement space of interferometric data, the incompletely sampled Fourier $(u,v)$ plane. After defining interferometric visibilities and their noise properties, we provide detailed instruction and examples for each of the steps in the method: 
    \begin{enumerate} 
        \item Generation of a {$(u,v)$ plane} filter which approximates the true emission pattern. 
        \item Cross-correlation of this filter with the measured visibilities. 
        \item Spectrum normalization and detection inference. 
        \item Line stacking (where applicable). 
    \end{enumerate} 
    
    \subsection{Matched Filtering}
        The matched filter can be derived in a number of ways. Here, we introduce its derivation by maximizing the SNR of a signal, but it can equivalently be interpreted as a least squares estimator \citep[see e.g.][]{Schwartz_1975, Vio_2016}. In general, a signal $\bm{s}$ may be corrupted by additive white noise $\bm{v}$, yielding an observation $\bm{x} = \bm{s} + \bm{v}$. To maximize the SNR of this signal by applying a linear filter $\bm{h}$, we can first write the SNR (using the definition of signal power/noise power) as
        \begin{equation}
            \mathrm{SNR} = \frac{\bm{h}^{*}\bm{s}\bm{s}^{*}\bm{h}}{\bm{h}^{*}\bm{R_{v}}\bm{h}},
        \end{equation}
        where $^{*}$ denotes the conjugate transpose and $\bm{R_v} = E[\bm{v}\bm{v}^{*}]$ is a covariance matrix of the noise $\bm{v}$, where $E[~]$ is the expectation operator. Under these conditions, the filter $\bm{h}$ which maximizes SNR is
        \begin{equation}
            \bm{h} = [\frac{1}{\sqrt{\bm{s}^{*}\bm{R_{v}}^{-1}\bm{s}}}]\bm{R_{v}}^{-1}\bm{s} = C\bm{R_{v}}^{-1}\bm{s},
        \end{equation}
        i.e., the original signal $\bm{s}$ multiplied by the data weights $\bm{R_{v}}^{-1}$ and a normalization constant $C=1/\sqrt{\bm{s}^{*}\bm{R_{v}}^{-1}\bm{s}}$ \citep[e.g.]{Woodward_1953, North_1963, Cumming_2005}.
        
        A simple application of such a filter is locating a signal within a one dimensional dataset such as an emission spectrum. In this case, a short signal $\bm{s}$ of length $n_s$ is embedded within a longer noisy observed spectrum $\bm{x}$ of length $n_x$, with the location of $\bm{s}$ within $\bm{x}$ unknown. As long as the shape of $\bm{s}$ is known, a filter kernel $\bm{h}$ can be calculated using equation 2 and cross-correlated with $\bm{x}$ to locate $\bm{s}$. This cross-correlation is often thought of as a sliding dot product, and yields a one dimensional impulse response spectrum $\bm{T}$, of length $n_x-n_s+1$. Each element of $\bm{T}$ at a position $i_0$ will then be:
        \begin{equation}
            T_{i_0} = \sum_{i=i_0}^{i_0+N_s-1}x_{i}h_{i-i_0};~~~~i_0 \in [0, n_x-n_s].
        \end{equation}
        This impulse response spectrum loses physical significance that the original observed spectrum held (it is no longer in units of power or flux). It instead encodes the degree of similarity between the observations and the filter at any given point in the observed spectrum. By projecting the noisy observations in this way, the total noise is decreased and the SNR of the signal is increased. Because the filter is linear, the Gaussian nature of the noise is preserved. The impulse response spectrum can be easily examined for evidence of $\bm{s}$, with a detection threshold set to some multiple of the standard deviation of $\bm{T}$ (e.g. 4$\sigma$), or a false positive rate scaled with the variance of $\bm{T}$.

    \subsection{Interferometric Visibilities}
        Each interferometric visibility, $V_i$, is measured as the complex product of the output of the first antenna of a baseline pair in the array with the complex conjugate of the output of the second antenna \citep[see e.g.][]{Thompson_2017}. The projected baseline distance between the two antennas then defines the location of the visibility on the $(u,v)$ plane. Each visibility $V_i$ is associated with a unique weight $w_i=1/\sigma_i^2$, where $\sigma_i^2$ encodes the variance of $V_i$ (see Appendix \ref{appendix:C} for more details on data weights in interferometric datasets). In addition to being measured at discrete spatial frequencies, the visibilities are also measured at a series of spectral frequencies (channels). Cross-correlation in this discretely sampled three dimensional $(u,v,channel)$ space is computationally awkward, but the dataset can be reshaped to a two-dimensional dataset of size $(n_{uv}, n_{c})$, where each visibility row corresponds to a unique location on the $(u,v)$ plane in units of distance. Both the complex visibilities and their corresponding real weights are stored this way in the UVFITS\footnote{The UVFITS format definition can be found in AIPS Memo \#114 at \url{http://www.aips.nrao.edu/aipsmemo.html}} and Measurement Set (MS)\footnote{The MS format definition can be found at \url{https://casa.nrao.edu/Memos/229.html}} formats of the Common Astronomy Software Applications package (CASA).
    
        Transforming between visibility space and image space requires a gridding and deconvolution routine, such as \texttt{CLEAN}, in one direction and a visibility sampling routine in the other direction, such as \texttt{uvmodel} in MIRIAD, or \texttt{simobserve} in CASA. As using the full \texttt{simobserve} task is relatively slow and \texttt{uvmodel} is not easily interfaced with Python, we have written a Python based visibility sampling routine, \texttt{vis$\textunderscore$sample}, which is able to interface with CASA MS and UVFITS formats. This package builds on an implementation of the sampling algorithm in the DiskJockey package \citep{Czekala_2015, Czekala_2016} identical to that used in \texttt{uvmodel} and \texttt{simobserve} and uses the spheroidal gridding function approximations described by \cite{Schwab_1984}. As identical algorithms and gridding functions are used, output from \texttt{vis$\textunderscore$sample} is identical to output from \texttt{uvmodel} and \texttt{simobserve}. \footnote{In addition to its utility for filter kernel generation, we note that \texttt{vis$\textunderscore$sample} may be useful for visibility fitting of modern interferometric datasets \citep[e.g.][]{Macgregor_2016, Loomis_2017}. \texttt{vis$\textunderscore$sample} is publicly available under the MIT license at \url{https://github.com/AstroChem/vis$\textunderscore$sample} or in the Anaconda Cloud at \url{https://anaconda.org/rloomis/vis$\textunderscore$sample}}

    \subsection{Filter Kernel Generation}
        The principle assumption of a matched filter analysis is that the shape of the signal $\bm{s}$ is known, or can be reasonably approximated. In traditional applications, such as RADAR, the outbound signal is user-generated and therefore the exact form is known. In astronomical applications, however, the ideal matched filter kernel is unknown and must be approximated. As it is unknown how closely the filter approximates the true signal, any derived detection significance will be a lower limit. The method is relatively robust to choice of filter, however, as long as the filter is a reasonable approximation of the source spatio-kinematic structure.
        
\begin{figure*}[ht!]
        \centering
        \includegraphics[width=0.95\textwidth]{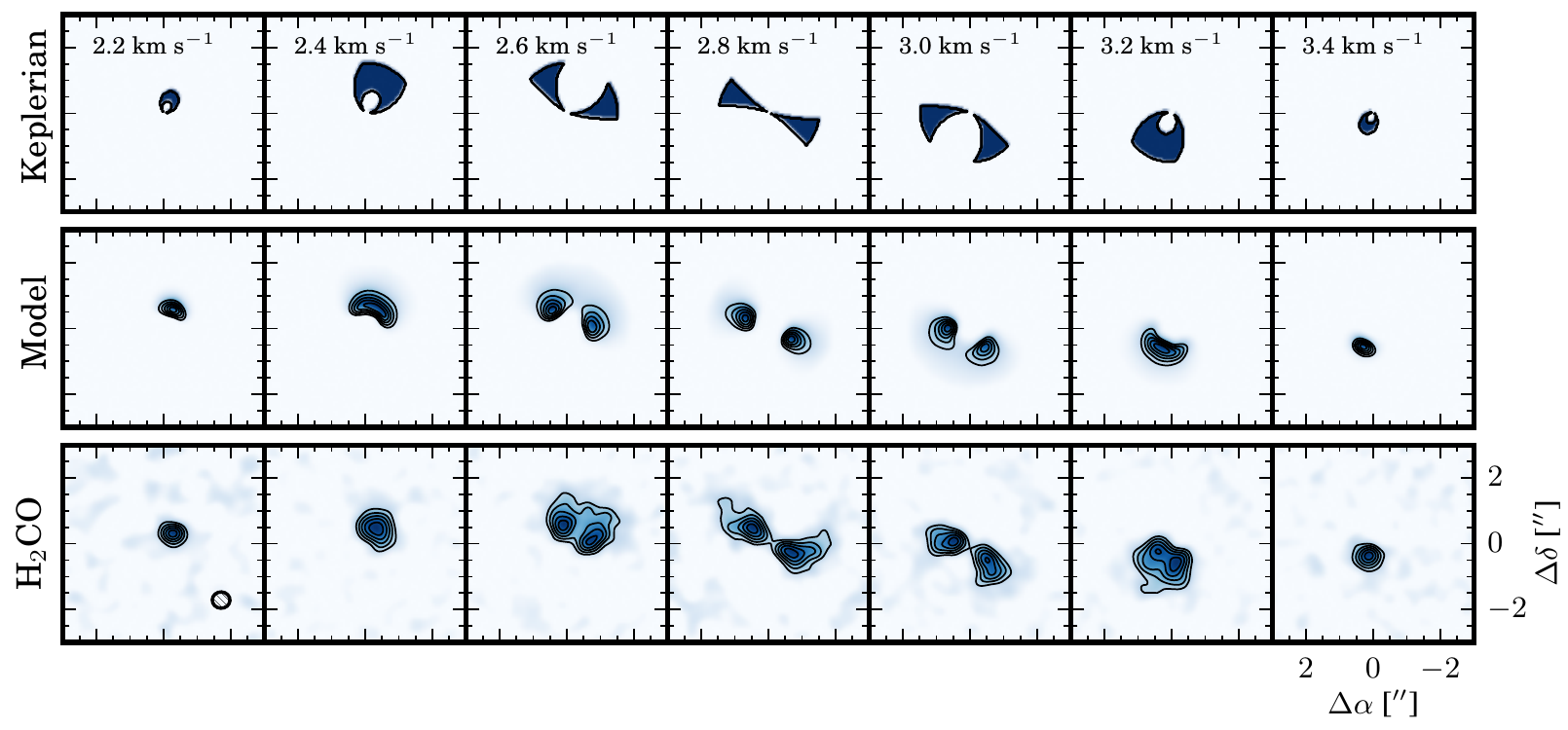}
        \caption{{\small Three examples of channel maps used to generate filter kernels for emission from the protoplanetary disk around TW Hya. \textit{Top:} a simple kernel based on Keplerian rotation. \textit{Middle:} a kernel based on a parametric model of CH$_3$OH from \cite{Walsh_2016}. \textit{Bottom:} a data-driven kernel generated from H$_2$CO observations from \cite{Oberg_2017}. All kernels have 0.2~km~s$^{-1}$ channels and are normalized by their peak intensities.} \label{Figure 2}}
        \end{figure*}

        We suggest two possible approaches: (1) calculating a kernel from a model of the source (model-driven), or (2) calculating a kernel from prior observations of strong emission lines (data-driven). In both cases the kernel is first constructed in the image plane and then Fourier transformed and visibility sampled to match the $(u,v)$ coverage of the observations. The approximated signal $\bm{f}$ and the inverted noise covariance matrix $\bm{R_v^{-1}}$ (calculated from the observational data weights, see Appendix \ref{appendix:C}) are then used to compute the full filter kernel, including the normalization prefactor:
        \begin{equation}
            \bm{h} = [\frac{1}{\sqrt{\bm{f}^{*}\bm{R_{v}}^{-1}\bm{f}}}]\bm{R_{v}}^{-1}\bm{f}.
        \end{equation}
        Fig. \ref{Figure 2} presents three examples of the different filter kernel estimation approaches. First, for objects such as protoplanetary disks or galaxies, the source inclination and position angle are often well-known and a spatio-kinematic model of the gas can be approximated. In the top panels of Fig. \ref{Figure 2} we have generated a Keplerian mask for molecular emission from the protoplanetary disk around TW Hya. Alternatively, a more detailed filter kernel can be generated from an astrochemical model of the source, with emission calculated using a radiative transfer code such as \texttt{RADMC-3D} \citep{Dullemond_2012} or \texttt{LIME} \citep{Brinch_2010}. An example is shown in the middle panels of Fig. \ref{Figure 2}, generated from the parametric CH$_3$OH abundance model in \cite{Walsh_2016}. Details of the model are presented in Appendix \ref{appendix:D}.
        
        In the data-driven approach, an assumption is made that an observed molecular transition shares its spatio-kinematic pattern with the desired weak line. The filter will be most effective when the template and target lines have well-matched spatial distributions, e.g. if the two lines are a strong and a weak line, respectively, of the same molecule. In \cite{Carney_2017}, we used this approach to detect weak H$_2$CO lines in HD 163296, using a stronger H$_2$CO line as a data-driven filter (see \S4.1 and \S4.4). Similarly, lines of a known species can be used as a filter for an undetected but chemically related molecule that is presumed to be co-spatial. The bottom row panels of Fig. \ref{Figure 2} present observations of H$_2$CO around TW Hya \citep{Oberg_2017} which could be used as a filter for CH$_3$OH emission, due to their linked formation pathways \citep[e.g.,][]{Cuppen_2009, Qi_2013, Walsh_2014, Loomis_2015}. Details of the observations and kernel generation are presented in Appendix \ref{appendix:D}.
        %The simplest implementation of this approach is to use visibilities from an observation of a known line directly as the template filter kernel. This rarely optimal, however, as the individual visibilities are often noise-dominated and may not be sampled at the same spectral resolution as the weak line data. A better approach involves an intermediate step of generating an image cube from the data using a deconvolution routine such as \texttt{CLEAN}. The image cube is noise clipped and then smoothed to yield a signal-dominated approximation of the true emission distribution. This image cube can then be interpolated to the velocity resolution of the target line data and transformed back into a (mostly) noiseless set of kernel visibilities through a 
        
    \subsection{Computing the Impulse Response Function}    
        \begin{figure*}[ht!]
        \centering
        \includegraphics[width=0.83\textwidth]{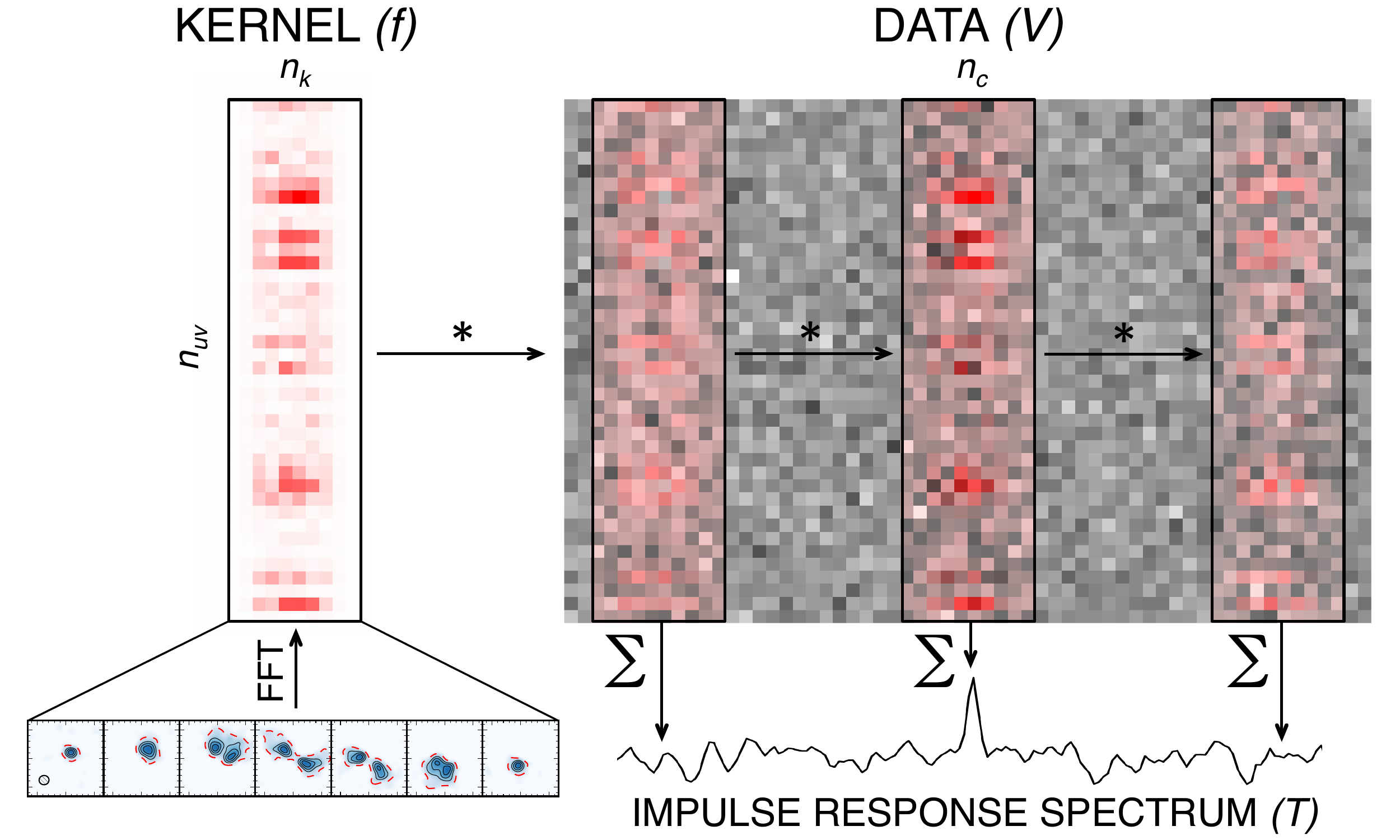}
        \caption{{\small A diagram illustrating how the filter kernel is cross-correlated with the data to produce a spectrum of the impulse response to the filter. The kernel is shown on the left, with dimensions of $n_{k}$ channels horizontally and $n_{uv}$ visibilities vertically (not shown to scale). Several representative channels of the kernel are shown imaged. The amplitudes of the complex kernel have been binned and pixelated to be visually intuitive. The complex data shown in gray-scale is also binned and pixelated, and has an identical number of visibilities, but $n_{c} \gg n_{k}$. The filter is applied to the data as a sliding inner product, and three illustrative regions are shown to visualize the cross-correlation at various points. Within these regions, a stronger red color signifies a stronger correlation with the corresponding kernel value, and the real portion of the response is summed over the entire region to produce the corresponding impulse response for each channel, with the line detected in the central channel.} 
        \label{Figure 3}}
        \end{figure*}
    
        Fig. \ref{Figure 3} schematically diagrams how these filter kernels would be applied to the data to produce an impulse response spectrum. First, the image plane kernel is Fourier transformed and visibility sampled to produce a complex $(u,v)$ plane kernel $\bm{f}$ of size $(n_{uv}, n_{k})$. The inverted noise covariance matrix $\bm{R_v^{-1}}$ is calculated from the data weights $\bm{w}$ and combined with $\bm{f}$ using Eq. 4 to produce the full filter kernel $\bm{h}$. This kernel is then cross-correlated with the data $\bm{V}$, of size $(n_{uv}, n_{c})$. The kernel and the data both have the same number of visibilities, $n_{uv}$, but different numbers of channels, and the kernel slides through the data along the spectral axis. At each channel, the filter impulse response spectrum $\bm{T}$ is calculated by taking the complex inner product of the windowed data with the kernel:
        \begin{equation}
            T_{i_0} = \sum_{i=i_0}^{i_0+n_{k}-1}\sum_{j=0}^{n_{uv}}V_{i,j}h_{i-i_0,j};~~~i_0 \in [0, n_c-n_k].
        \end{equation}
        As the data is complex, the filter response spectrum is also complex with a normalized total noise power. Signal power will leak from the real to the imaginary portion of the response, however, if there is a phase misalignment between the sky locations of the filter and the source. Thus if the filter has been properly phase shifted to be aligned with the source, the resultant impulse response spectrum $\bm{T}$ can be written as:
        \begin{equation}
            T_{i_0} = \mathrm{Re}\bigg[\sqrt{2}\sum_{i=i_0}^{i_0+n_{k}-1}\sum_{j=0}^{n_{uv}}V_{i,j}h_{i-i_0,j}\bigg];~~~i_0 \in [0, n_c-n_k],
        \end{equation}
        with the factor of $\sqrt{2}$ introduced to normalize the noise power in the real portion of the spectrum.
        
        This method of calculating the cross-correlation is conceptually simple, but computationally inefficient. Computing inner products of the windowed data requires either manipulation of the (very large) dataset in memory or non-sequential memory access, preventing speed increases through vectorization.\footnote{Using FFT cross-correlation is even slower for typical interferometric dataset sizes.} There is no restriction, however, on the order of operations in which the inner products are internally calculated. We use this to our advantage and treat the partial two-dimensional cross-correlation as a series of $n_{uv}$ one dimensional cross-correlations along the spectral axis, yielding $n_{uv}$ individual impulse response curves. The UVFITS and MS data formats store visibilities in a row-major order such that these one-dimensional cross-correlations quickly access data sequentially in memory. The resulting impulse response curves are then summed along the spatial frequency dimension, identical to Eq. 6, but with the order of the summations switched,
        \begin{equation}
            T_{i_0} = \mathrm{Re}\bigg[\sqrt{2}\sum_{j=0}^{n_{uv}}\sum_{i=i_0}^{i_0+n_{k}-1}V_{i,j}h_{i-i_0,j}\bigg];~~~i_0 \in [0, n_c-n_k],
        \end{equation}
        yielding a final impulse response spectrum identical to that from the sliding window method shown in Fig. \ref{Figure 3}. The $n_{uv}$ one-dimensional cross-correlations are independent, making parallelization trivial. Using this approach, the full bandwidth of a typical ALMA spectral window (e.g. 3840 channels over a 1~hr integration with 43 antennas) can be filtered very quickly on a desktop (e.g. a few seconds on a quad-core 3.3GHz processor).

        %As the filter response shown in Figure 2 is a slice of the full two-dimensional convolution of the kernel and the data, one might imagine utilizing fast two-dimensional convolution techniques. Two-dimensional convolutions are generally calculated either through Fourier transform methods or brute-force computation of all windowed dot products. The latter method is identical to the sliding dot product already described, and offers no speed increase, while the former leverages the convolution theorem of the Fourier transform. Fourier techniques offer large speed increases in situations where the kernel size is very small compared to the dimensions of the data, such as when applying a small Gaussian blur kernel to an image, but become slow when the size of the kernel and the data are comparable. In our case, both the data and the kernel share a size n$_{vis}$ in one of the dimensions, making Fourier methods of calculating the convolution ineffective.
        
    \subsection{Normalization and Detection Inference}
        Assessing the probability of a line detection from the filter impulse response requires understanding the noise properties of the response spectrum, which no longer holds the same physical significance as an emission spectrum. The response spectrum at a given frequency now represents how closely the data correspond to the filter, rather than the flux at that frequency. As the filter is linear, uncorrelated thermal noise in the visibilities manifests as Gaussian noise in the filter response. If the data weights are properly calibrated (see Appendix \ref{appendix:C}), the prefactor in Eq. 4 will normalize the filter response such that the spectrum has units of standard deviations ($\sigma$) with a RMS noise level of unity.
        
        In practice, however, we have found that the calculated data weights are often only accurate to $\sim$20\% compared to the actual variance of the visibilities. Thus we strongly recommend comparing the data weights and visibility scatter and recalculating the weights using a task such as \texttt{statwt} in CASA if there is a discrepancy. Alternatively, the filter response itself can be manually normalized by dividing by the standard deviation of the spectrum (excluding any channels with obvious signal).
        
        Additionally, the linear nature of the filter means that any unsubtracted continuum emission will result in a constant offset in the response spectrum. A model of the continuum can either be subtracted in the $(u,v)$ plane as a pre-processing step prior to filtering using a task such as \texttt{uvcontsub} in CASA, or be subtracted after filtering by subtracting the mean of signal-free channels in the impulse response spectrum. Subtraction after filtering removes the possibility of small inaccuracies in the $(u,v)$ subtracted model, to which the filter will be sensitive. Conversely, subtraction prior to filtering may be more convenient when a large bandwidth is covered for a source with a non-zero spectral index.
        
        Once the response spectrum is normalized and any offsets are removed, peaks can then be evaluated against a detection threshold, set at some number of standard deviations corresponding to an acceptable false alarm rate. We stress, however, that the detection significance is a lower limit, as it is unknown how closely the filter approximates the ideal matched filter.
        
        %Additionally, if multiple filters are tested on the data, the detection statistic must be corrected for multiple hypothesis testing after assessing the degree of independence between the different filters. More details about false alarm rates in matched filtering can be found in \cite{Vio_2016} and \cite{Vio_2017}.

    \subsection{Comparison to Image-Plane Spectral Extraction}
        As a proof of concept, we apply the method to synthetic observations of CH$_3$OH emission in TW Hya and compare to an aperture-based spectral extraction in the image plane. The modeled emission from the middle panels of Fig. \ref{Figure 2} is used to generate both the observations (with noise added) and the filter kernel. As this is a true matched filter, it provides a useful benchmark for comparison with the approximate filter results as presented in \S3.
        
        \begin{figure*}[ht!]
        \centering
        \includegraphics[width=0.8\textwidth]{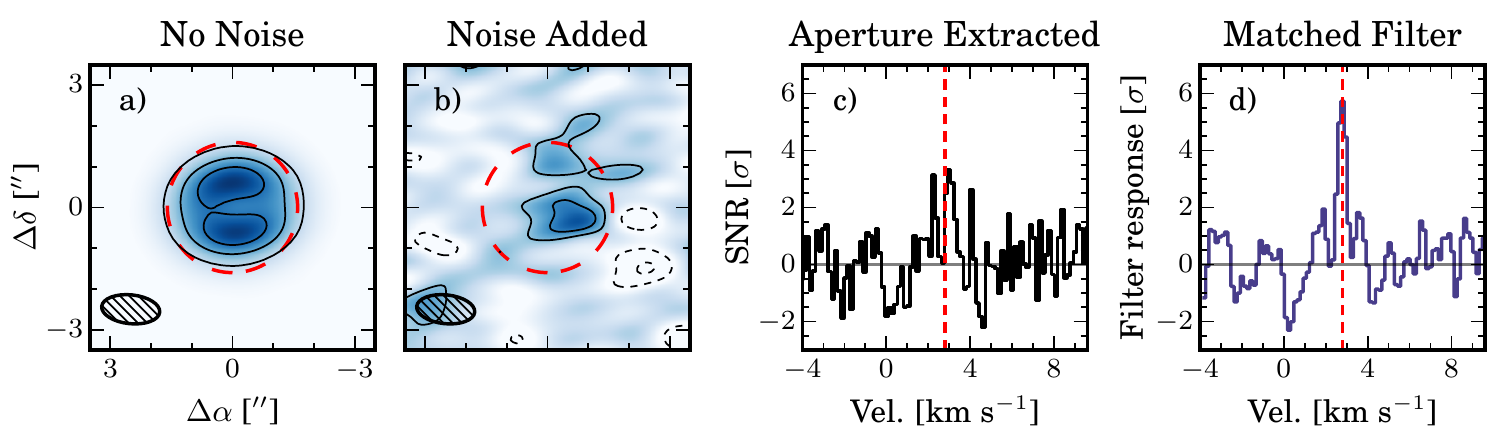}
        \caption{{\small Comparison of the ideal matched filter with conventional spectral extraction through aperture masking. \textit{Panel a:} Moment-0 map of simulated, noiseless CH$_3$OH 3$_{12}$-3$_{03}$ emission. The synthesized beam is shown in the lower left. Contours are [-3, -1.5, 1.5, 3]$\times$3.3 mJy bm$^{-1}$ km s$^{-1}$, corresponding to 1$\sigma$ in panel b. \textit{Panel b:} Moment-0 map of simulated and noise-corrupted CH$_3$OH emission. \textit{Panel c:} Spectrum of the noise-corrupted emission, extracted using an aperture 3$\arcsec$ in diameter. \textit{Panel d:} Ideal matched filter response to the noisy emission, with units of $\sigma$.} \label{Figure 4}}
        \end{figure*}
            
        A synthetic measurement set of observations was created from the CH$_3$OH emission cube described in \S2.3 by visibility sampling at baselines corresponding to the observations from \cite{Walsh_2016} using \texttt{vis$\textunderscore$sample}. The complex visibilities were then noise corrupted such that the rms noise was 5 mJy bm$^{-1}$ across each 0.15 km/s channel, equivalent to the \cite{Walsh_2016} observations. The noiseless and noisy measurement sets were imaged in CASA using the \texttt{CLEAN} task, with a \texttt{CLEAN} mask generated from the \texttt{LIME} output emission profile and a circular 1$\arcsec$ FWHM Gaussian taper applied in the Fourier plane to increase the SNR of the images. Only the noiseless measurement set was \texttt{CLEAN}ed; the noisy measurement set was dirty imaged to prevent bias from over-\texttt{CLEAN}ing, as the emission is practically at the noise limit in any given channel. Integrated intensity (moment-0) maps of the noiseless and noisy 3$_{12}$-3$_{03}$ transitions are shown in Fig. \ref{Figure 4}, panels a and b, respectively, and were generated by integrating all channels with emission. No clipping threshold was used. In the noisy case, the moment-0 map rms is $\sim$3.3 mJy bm$^{-1}$ km s$^{-1}$ and the peak integrated flux is $\sim$13.2 mJy bm$^{-1}$ km s$^{-1}$, yielding a SNR of $\sim$4. A spectrum was extracted from the noisy image cube using an aperture 3$\arcsec$ in diameter, equivalent to the extent of the CH$_3$OH emission (Fig. \ref{Figure 4}, panel c). The spectrum has a peak flux of $\sim$11.4 mJy and a rms noise of $\sim$3.2 mJy, yielding a SNR of $\sim$3.5$\sigma$. The rms noise level of the noisy spectrum was estimated from all channels without significant emission (i.e., excluding a velocity range of $\pm$ 1.5 km s$^{-1}$ around the systemic velocity of 2.8 km s$^{-1}$).
        
        We cross-correlate the CH$_3$OH filter kernel with the synthetic observations, as described in \S2.4, generating the filter response shown in Fig. \ref{Figure 4} panel d. The peak value of the filter response, 5.7$\sigma$, is the maximum SNR extractable from the data and represents a $\sim$40$\%$ and $\sim$60$\%$ improvement over the moment-0 and spectral detections, respectively. This already corresponds to a factor of 2--3 increase in effective observing time, but as discussed in both \S4.1 and Appendices \ref{appendix:A} \& \ref{appendix:B}, the level of possible SNR improvements will be higher for data sets that are better-resolved.
        
    \subsection{Stacking}
        Spectral stacking is a common method of SNR improvement for observations of multiple transitions of the same molecule \citep[e.g.][]{Langston_2007, Kalenskii_2010, Loomis_2016, Walsh_2016}. If the excitation conditions of two or more transitions are similar and their rest frequencies are well known, then the signals can be combined through weighted averaging,
        \begin{equation}
            \bm{T_s} = \sum_{i=0}^{n_s} \bm{T_{i}}w_{i},
        \end{equation}
        where the stacked spectrum $\bm{T_s}$ is generated by summing $n_s$ individual spectra $\bm{T_i}$ multiplied by weights $w_i$, proportional to the SNR of each $\bm{T_i}$. Knowledge of the relative strengths of each transition is therefore important to gain the most signal improvement. Application of a matched filter results in an estimated SNR for each transition, which can be used as a proxy for their relative strengths. The resultant impulse response spectra are then easily stacked to generate an appropriately weighted stacked spectrum.

        \begin{figure}[ht!]
        \centering
        \includegraphics[width=\columnwidth]{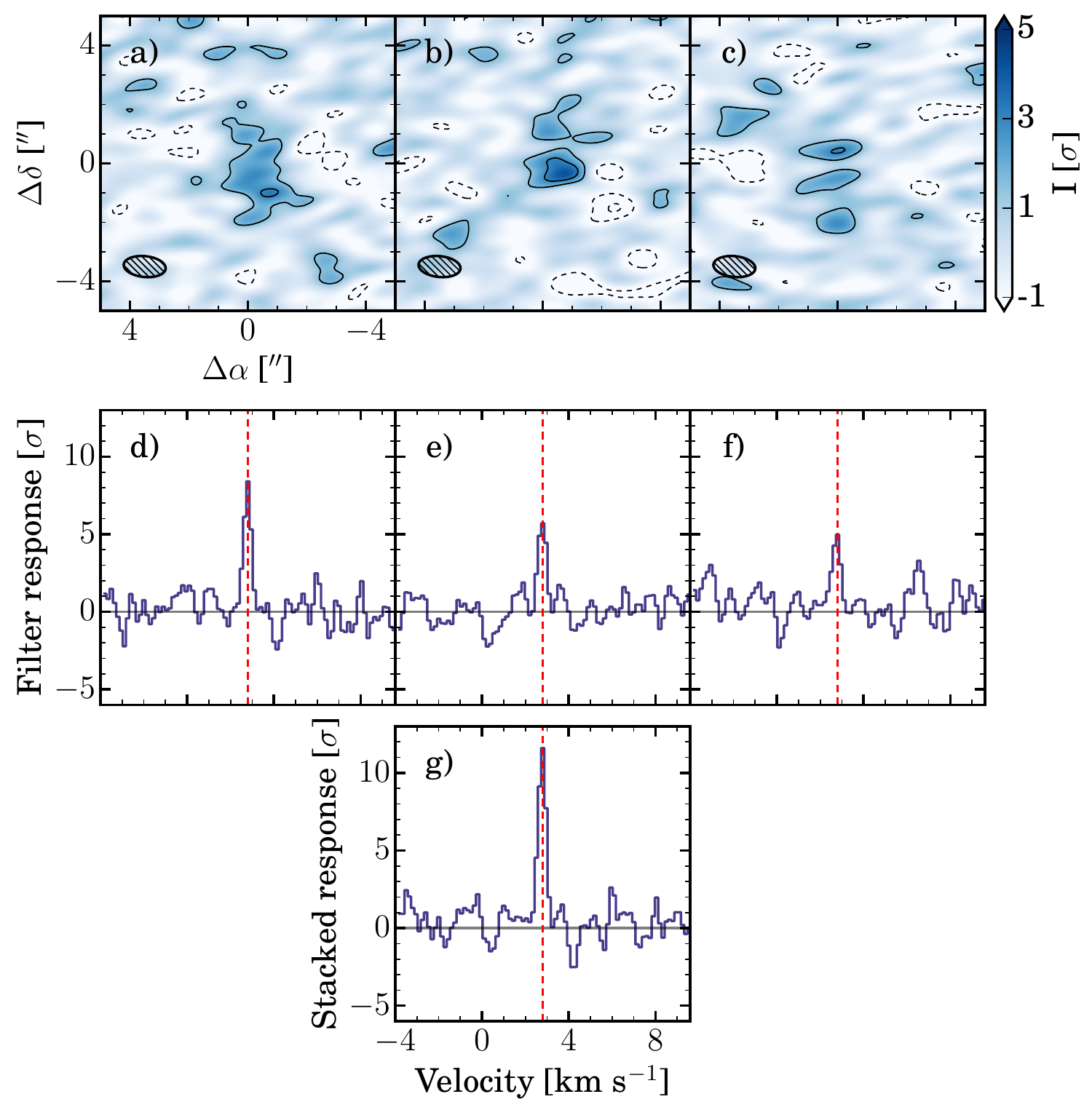}
        \caption{{\small Demonstration of line stacking on synthetic CH$_3$OH emission. \textit{Panel a-c:} Moment-0 maps of the three simulated and noise-corrupted transitions, 2$_{11}$-2$_{02}$, 3$_{12}$-3$_{03}$, and 4$_{13}$-4$_{04}$. Contours are [-3, -1.5, 1.5, 3]$\times\sigma$, $\sigma$=3.3 mJy bm$^{-1}$ km s$^{-1}$. The synthesized beam is shown in the lower left. \textit{Panel d-f:} Ideal matched filter response spectra. Peak SNRs are 8.4, 5.7, and 4.9$\sigma$. \textit{Panel g:} Filter response spectrum created by stacking the individual spectra from panels d-f. Each spectrum was weighted by its SNR, and the resultant spectrum has a SNR of 11.6.} \label{Figure 5}}
        \end{figure}

        To illustrate this process, we have repeated the simulations and filtering described in \S2.6 for three CH$_3$OH transitions: 2$_{11}$-2$_{02}$, 3$_{12}$-3$_{03}$, and 4$_{13}$-4$_{04}$, with relative strengths of 1.8:1.3:1.0. Moment-0 maps of the emission from each of these transitions are shown in Fig. \ref{Figure 5}, panels a, b, and c, with peak integrated fluxes of 11.7, 13.2, and 8.5 mJy km s$^{-1}$ and corresponding SNRs of 3.5, 4, and 2.6$\sigma$, respectively. The individual filter responses are shown in Fig. \ref{Figure 5}, panels d, e, and f, with peak SNRs of 8.4, 5.7, and 4.9$\sigma$, respectively. The filter responses were stacked using a weighted average, yielding the spectrum shown in Fig. \ref{Figure 5}, panel g, with a peak SNR of 11.6$\sigma$. The ratio of the filter responses $\sim$(1.7:1.2:1) recovers the flux ratio of the input models (1.8:1.3:1.0) fairly well, even though the 2$_{11}$-2$_{02}$ transition appears weaker than would be expected in the imaged data (likely due to random noise fluctuations in the inherently more noisy moment-0 maps). This highlights one of the advantages of applying the matched filter in the Fourier plane.

    %    The detection of \cite{Walsh_2016} illustrates that with low abundance species such as CH$_3$OH, individual lines are likely to be weak and only marginally detected, if at all. Chemical disk models suggest, however, that emission from multiple lines of the same species are likely to be highly correlated spatially, and extremely well correlated in their velocity patterns (e.g. Walsh et al. 2015). Additionally, most  are asymmetric tops and have a 'ladder' of lines, enabling multiple transitions to be observed in a single correlator setup with modern interferometers (CITATIONS). The matched filter technique is then ideally suited to this noise-dominated broadband regime, allowing for fast searches of multiple COMs by stacking their filter responses.

\section{Application to real ALMA data}
    \begin{figure*}[ht!]
    \centering
    \includegraphics[width=0.7\textwidth]{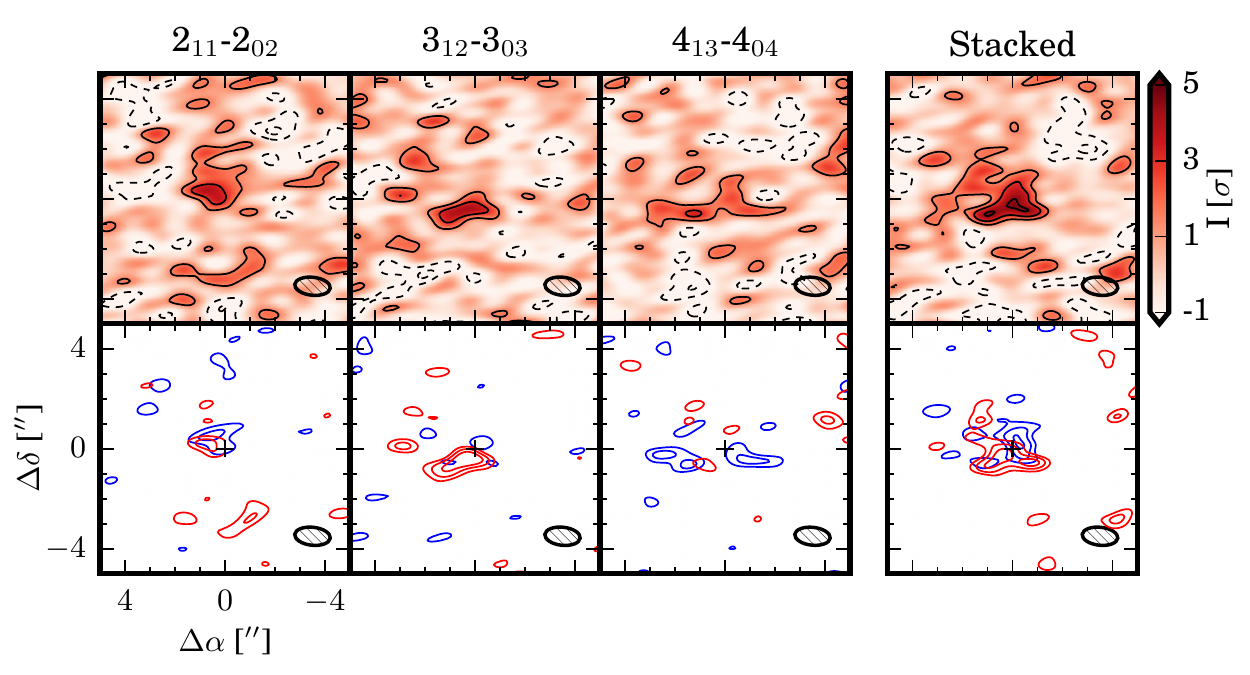}
    \caption{{\small CH$_3$OH observations toward TW Hya. \textit{Top:} CH$_3$OH emission from the 2$_{11}$-2$_{02}$, 3$_{12}$-3$_{03}$, and 4$_{13}$-4$_{04}$ transitions, and all three stacked. Contours are [-3, -1.5, 1.5, 3, 4.5]$\times\sigma$, $\sigma$=$\sim$3.6 mJy bm$^{-1}$ km s$^{-1}$ for the individual transitions and $\sim$2.3 mJy bm$^{-1}$ km s$^{-1}$ for the stacked image. \textit{Bottom:} same as top, but for 1 km s$^{-1}$ velocity bins around the source velocity, showing the disk rotation.} 
    \label{Figure 6} }
    \end{figure*}
    
    Matched filtering provides clear benefits when the ideal filter kernel is known. To explore its utility when the filter is approximated, we have applied the method to real ALMA Band 7 observations of CH$_3$OH toward TW Hya (project 2013.1.00902.S, P.I. C. Walsh), using all three kernels shown in Fig. \ref{Figure 2}. Details of the CH$_3$OH observations are presented by \cite{Walsh_2016}. They reported that the three observed CH$_3$OH transitions (2$_{11}$-2$_{02}$, 3$_{12}$-3$_{03}$, and 4$_{13}$-4$_{04}$) were not conclusively detected in any of the individual data cubes, and therefore only presented the stacked imaging data with a 5.1$\sigma$ detection in an aperture extracted spectrum. Moment-0 maps of the three observed CH$_3$OH transitions are presented in Fig. \ref{Figure 6}, with peak SNRs of 4.3, 4.3, and 2.9$\sigma$, respectively. A stacked moment-0 map is shown on the far right with a peak SNR of 4.8$\sigma$. The lower set of panels in Fig. \ref{Figure 6} show binned red and blue-shifted emission, highlighting the disk rotation. Rotation about the principal axis is seen for the stacked emission, and hinted at for two of the individual transitions.

    \begin{figure}[ht!]
    \centering
    \includegraphics[width=\columnwidth]{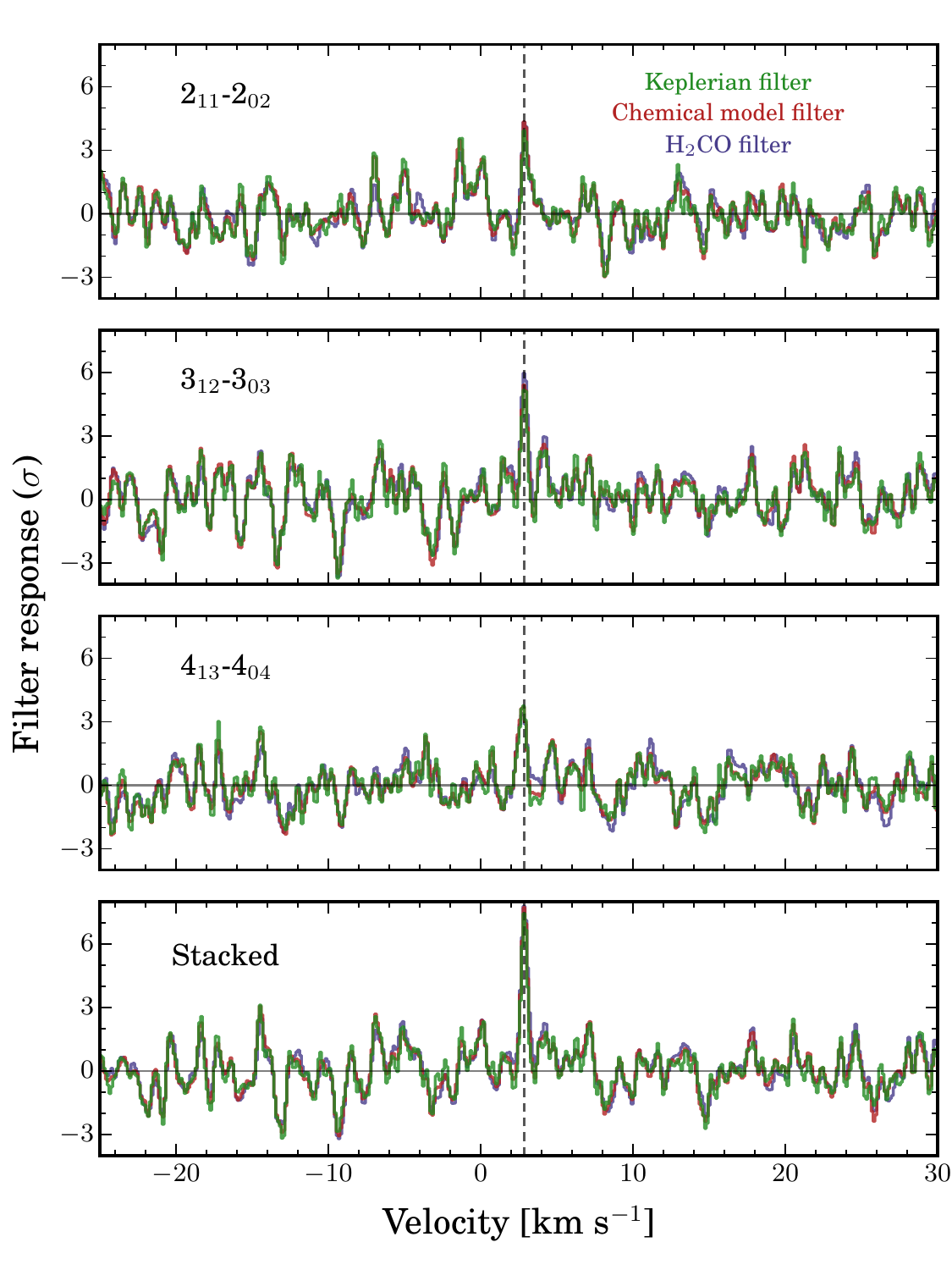}
    \caption{{\small Filter response spectra for each CH$_3$OH transition. The impulse responses to the Keplerian, CH$_3$OH model, and H$_2$CO filters are shown in green, red, and blue, respectively.} \label{Figure 7}}
    \end{figure}

    %Slight differences in the final stacked SNR, however, may provide further insight to the CH$_3$OH distribution. The simple Keplerian filter performs poorest, likely due to an inaccurate emission morphology, as line broadening and radially decreasing column density and temperature were not taken into account. The CH$_3$OH model and H$_2$CO data-driven filters perform better, with

    Each of the three filter kernels from Fig. \ref{Figure 2} were cross-correlated with the visibilities of each of the observed CH$_3$OH transitions, producing the filter responses shown in Fig. \ref{Figure 7}. All three filters detect the individual lines and show similar SNR boosts, demonstrating that the method is robust to the choice of filter. The H$_2$CO filter yields the strongest responses for the individual lines (4.3, 6.0, and 3.4$\sigma$, respectively). Stacking these spectra together, the H$_2$CO filter yields a robust detection of 7.8$\sigma$, a 53$\%$ improvement over the 5.1$\sigma$ detection reported in \cite{Walsh_2016}. The Keplerian and CH$_3$OH model filters produce stacked responses of 7.4$\sigma$ and 7.7$\sigma$, respectively.
    
    %As the data responds more strongly to the H$_2$CO filter than the CH$_3$OH model filter, we can infer that the true CH$_3$OH distribution is likely more similar to H$_2$CO than the model, perhaps due to a smaller inner radius of the CH$_3$OH emission ring than was reported from the model fit in \cite{Walsh_2016}, where a monotonically decreasing column density with radius was enforced. Higher resolution observations of the CH$_3$OH emission at higher SNR will be able to probe the CH$_3$OH inner radius and test this prediction.
     
    \begin{figure*}[ht!]
    \centering
    \includegraphics[width=0.8\textwidth]{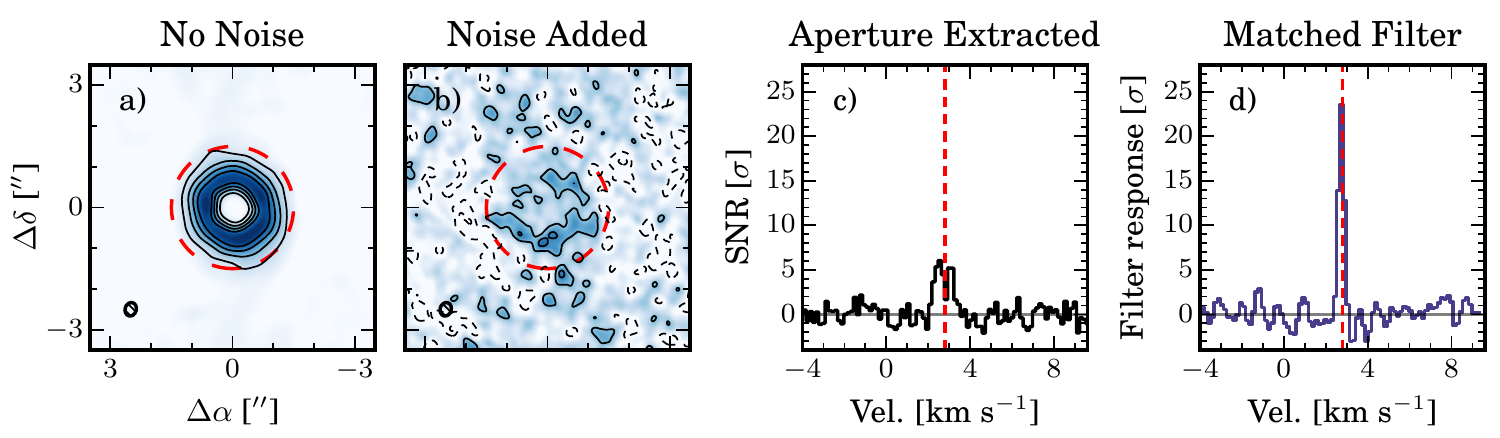}
    \caption{{\small Comparison of matched filtering with traditional methods, as in Fig. \ref{Figure 4} but at higher spatial resolution. \textit{Panel a:} Moment-0 map of simulated noiseless CH$_3$OH 3$_{12}$-3$_{03}$ emission. The synthesized beam is shown in the lower left. Contours are [-3, -1.5, 1.5, 3]$\times$5.6 mJy bm$^{-1}$ km s$^{-1}$, corresponding to 1$\sigma$ in panel b. \textit{Panel b:} Moment-0 map of simulated and noise-corrupted CH$_3$OH emission. \textit{Panel c:} Spectrum of the noise-corrupted emission, extracted using an aperture 3$\arcsec$ in diameter. \textit{Panel d:} Ideal matched filter response to the noisy emission. Units are $\sigma$, defined as the rms filter response in channels with no emission.} \label{Figure 8}}
    \end{figure*}  
       
\section{Discussion}
    We have presented a formulation of matched filtering for interferometric spectral line data and shown that this technique can improve SNR and therefore line detectability in both synthetic and real test cases. We now discuss how much of a SNR boost one might expect for a given dataset, compare to alternative techniques, and suggest potential further applications of this method.

    \subsection{Factors Affecting SNR Boost}
        Compared with traditional line detection methods, the matched filter approach offers an improved SNR. The degree of SNR boost depends on both the accuracy of the approximated kernel as well as the specific properties of the data (particularly the spatial resolution). In the synthetic and real data test cases presented in \S3, we found that application of a matched filter could increase SNR by $\sim$60\%. The method was found to be similarly effective when applied to real data ($\sim$53\% vs $\sim$60\% boost), demonstrating that it is robust to the choice of approximated filter.
        
        Intuitively, the SNR boost and spatial resolution of the emission should be coupled. By definition, a spatially unresolved signal encodes no spatio-kinematic information, and in this limit the matched filter technique will provide no increase in SNR other than the boost from spectral averaging. As emission is spatially resolved, SNR will decrease roughly with the square of the degree of spatial resolution (source width / beam size), with additional losses due to spatial filtering \citep[see e.g.][]{Crane_1986}. With appropriate knowledge of the velocity structure, a matched filter essentially negates this effect, and thus the SNR boost scales directly with the spatial resolution of the signal \cite[see][for a detailed image-plane derivation of this SNR boost]{Yen_2016}. Fig. \ref{Figure 8} illustrates this effect, with an identical simulation to that shown in Fig. \ref{Figure 4}, but with a higher spatial resolution of $\sim$0$\farcs$3. The data were noise corrupted to reach a similar $\sim$4$\sigma$ detection in the moment map, although the SNR in the extracted spectrum is now $\sim$6$\sigma$, highlighting how ineffective moment maps are at high spatial resolutions. The filter response is also now much larger (SNR=23.6$\sigma$), yielding a SNR boost over the aperture extracted spectra of $\sim$400\%, compared to $\sim$60\% in Fig. \ref{Figure 4}. Similarly, application of matched filtering to higher resolution observations of H$_2$CO \citep{Carney_2017} produced a SNR gain of over 500\%, confirming in practice the relationship between SNR boost and spatial resolution.
    
    \subsection{Comparison to Other Methods}
        Recently \cite{Matra_2015} and \cite{Marino_2016} introduced an image-plane line detection technique \citep[also independently introduced and formalized by][]{Yen_2016} that provides some similar benefits to the matched filter technique. In their approaches, pixels from a dirty image are adjusted for an assumed velocity offset (from a source model), and the velocity corrected spectra are then stacked. In many ways, this can be seen as an image-plane analog to a Fourier plane matched filter, and it should yield comparable increases in SNR (see Appendix \ref{appendix:B} for more details). This is confirmed by comparing the results of the matched filtering technique to detect H$_2$CO in HD 163296 \citep{Carney_2017} with those obtained on the same dataset by \cite{Yen_2016}. In both cases, a SNR boost of $\sim$500\% is achieved.
        
        Several subtle differences between matched filtering and pixel stacking, however, may motivate their use in a synergistic fashion. First, application of a matched filter in the uv-plane requires no imaging of the data, and is therefore much faster and more robust than image-plane spectral stacking. Second, the matched filter technique allows for a more accurate emission model than simple Keplerian rotation to be applied to the data (i.e., the spatial distribution of molecular signal can be properly used for weighting). On the other hand, stacking pixels in the image plane makes extracting a flux measurement for the line much simpler and allows for a radial profile to be estimated \citep{Yen_2016}. The two techniques could therefore be used sequentially to exploit the benefits of each, with a matched filter first used to quickly identify and confirm line detections in a dataset and pixel stacking then used to better characterize the lines.
        
    \subsection{Line Flux Estimation}
        The main utility of matched filtering when applied to interferometric spectral line data is in the rapid detection of weak lines, rather than their detailed characterization. Once a line is identified, it might be further characterized through careful imaging, spectral stacking, or model-fitting to the visibilities. For the weakest lines, however, detailed characterization will likely require additional observations. Matched filtering provides useful predictive utility when planning these observations, robustly confirming weak lines which might be desirable targets.
        
        In particular, after a weak line is identified, the matched filter method can be used to estimate a line flux if the emission is too weak to be seen directly in the image cube. When a data-driven approach is taken, the responses of the target and template lines to the filter can be compared. If the two lines have a similar emission morphology, the ratio of their responses will be similar to the ratio of their fluxes, with the flux of the strong line being easy to measure. 
        
        This can be proven by considering a modified version of equation 1, writing down the SNR using the signal/rms definition:
        \begin{equation}
            \mathrm{SNR_s} = \sqrt{\frac{\bm{h}^{*}\bm{s}\bm{s}^{*}\bm{h}}{\bm{h}^{*}\bm{R_{v}}\bm{h}}}
        \end{equation}
        We can treat the two lines as signals $\bm{y}$ and $\bm{s}$, where $\bm{y}$ differs only from $\bm{s}$ by an arbitrary constant $\alpha$, i.e. $\bm{y} = \alpha \bm{s}$. The SNR of $\bm{y}$ after filter application is: 
        \begin{equation}
            \mathrm{SNR_y} = \sqrt{\frac{\bm{h}^{*}\alpha \bm{s}\alpha \bm{s}^{*}\bm{h}}{\bm{h}^{*}\bm{R_{v}}\bm{h}}}
        \end{equation}
        which reduces to:
        \begin{equation}
            \mathrm{SNR_y} = \alpha\sqrt{\frac{\bm{h}^{*}\bm{s}\bm{s}^{*}\bm{h}}{\bm{h}^{*}\bm{R_{v}}\bm{h}}}
        \end{equation}
        \begin{equation}
            \mathrm{SNR_y} = \alpha~\mathrm{SNR_s}
        \end{equation}
        Therefore we can use the filter impulse response ratio to roughly estimate the flux, with the accuracy dependent on the closeness of the filter kernel to the true emission distribution. It is important to note that this estimate will always be a lower limit. An upper limit can additionally be derived from the imaged weak line, bounding the flux measurement. This approach was used by \cite{Carney_2017} to determine the flux ratio of multiple detected H$_2$CO lines, enabling them to constrain the H$_2$CO excitation temperature.

    \subsection{Application to Line Surveys}
        In addition to aiding the detection of specific known weak lines, interferometric matched filtering provides substantial benefits for the processing of spectral line surveys where source locations and approximate spatio-kinematic structures are known. Imaging the full bandwidth of these large datasets at their native spectral resolution is a time consuming process, often taking many hours or even days. Because much of the information in these datasets is contained in spatio-kinematic patterns of the spectral lines, decreasing spectral resolution through channel averaging is typically not a viable option and can result in signal loss. A choice must therefore be made between imaging only small targeted windows of the broadband dataset, or spending time and computing resources on imaging the full bandwidth. For sparsely populated line surveys (e.g., of protoplanetary disks), imaging the entire data set is inherently inefficient, since most of the channels do not contain signal. Conversely, selective imaging reduces the likelihood of serendipitously detecting weak species, and conflicts with the motivations of an unbiased survey. 
        
        Numerous tools have been developed to aid in identifying spectral line emission in broadband datasets from current and future instruments such as ALMA, ASKAP, VLA, SKA, and the ngVLA \citep[e.g.,][]{Koribalski_2012, Whiting_2012a, Whiting_2012b, Friedel_2015, Serra_2015}, but these methods often rely on a fully imaged datacube as input. In instances where the locations of the sources being targeted are known a-priori, our presently described method of matched filtering can help streamline this process by quickly and robustly identifying lines in the native visibilities. Then only these lines need be imaged and analyzed. In sources with a single dominant velocity pattern, a strong line could be imaged, converted to a filter kernel, and cross-correlated through the entire dataset in a small fraction of the time it would take to image that same dataset. The resulting full-band impulse response spectrum then provides a convenient first look at the dataset, guiding the observer as to which sections of the data are worth windowing out for further imaging and analysis. In particular, matched filtering will highlight weak lines that the observer would have missed even in a careful \texttt{CLEAN} of the data.
        
        We note that our $(u,v)$-plane method could likely be extended to full 3D searches in blind surveys, where source locations are not known a-priori, but such an implementation is beyond the scope of this paper. It is not immediately clear whether the large speed benefits of $(u,v)$-plane analysis over full survey imaging would be maintained in such a 3D search space, and we encourage further research in this area.

\section{Conclusion}
    We have shown that the technique of matched filtering can be easily implemented for analyzing interferometric observations of spectral lines, significantly improving sensitivity when searching for weak lines. An open-source Python-based implementation is freely available under the MIT license at \url{https://github.com/AstroChem/VISIBLE}. As a case study, we have focused on observations of protoplanetary disks with ALMA, but our approach is applicable to spectral line data of any astronomical source with a spatio-kinematic pattern that can be used to generate a filter kernel, and will likely be beneficial for spectral line observations from a wide range of current and future instruments (e.g., the SKA and the ngVLA).

    We find that when applied to real data, the method results in large sensitivity increases, ranging from 53\% for CH$_{3}$OH in \cite{Walsh_2016}, to $\sim$500\% for H$_2$CO in \cite{Carney_2017}. The degree of sensitivity boost is proportional to the spatial resolution of the observations. These sensitivity increases are equivalent to factors of 2-25 in effective observing time, allowing observers to better leverage limited telescope resources. Additionally, the speed of the technique is beneficial when analyzing large bandwidth line surveys, robustly identifying all lines in a spectrum in a small fraction of the time it would take to image the same dataset. Finally, the method works synergistically with the methods presented in \cite{Matra_2015} and \cite{Yen_2016} and tools such as ADMIT, forming a comprehensive suite of analysis techniques for spectral lines in large interferometric datasets.

\acknowledgments
We would like to thank Andrew Vanderburg and Peter Williams for productive discussions and assistance with software package management and distribution. We also thank the anonymous referee for providing comments that greatly improved the quality of the manuscript. RAL and JH gratefully acknowledge funding from National Science Foundation Graduate Research Fellowships (Grant No. DGE-1144152). RAL also acknowledges funding from the NRAO Student Observing Support Program. KIO acknowledges funding from the Alfred P. Sloan Foundation and the David and Lucile Packard Foundation. CW acknowledges financial support from the Netherlands Organisation for Scientific Research (NWO, grant 639.041.335) and start-up funds from the University of Leeds, UK. The National Radio Astronomy Observatory is a facility of the National Science Foundation operated under cooperative agreement by Associated Universities, Inc.  This paper makes use of the following ALMA data: ADS/JAO.ALMA\#2013.1.00902.S and ADS/JAO.ALMA\#2013.1.00114.S. ALMA is a partnership of ESO (representing its member states), NSF (USA) and NINS (Japan), together with NRC (Canada) and NSC and ASIAA (Taiwan), in cooperation with the Republic of Chile. The Joint ALMA Observatory is operated by ESO, AUI/NRAO and NAOJ.

\software{Astropy \citep{Astropy_2013}, CASA \citep{McMullin_2007}, casa-python, DiskJockey \citep{Czekala_2015, Czekala_2016}, LIME \citep{Brinch_2010}, Matplotlib \citep{Hunter_2007}, NumPy \citep{Jones_2001}, SciPy \citep{vanderWalt_2011}, VISIBLE \citep{VISIBLE_Zenodo, VISIBLE_ASCL}}

\appendix

\section{Calculating SNR boost for a matched filter}
\label{appendix:A}
    SNR (using the definition of signal-power/noise-power) can be written for an arbitrary signal $\bm{s}$ and filter $\bm{h}$ as:
    \begin{equation}
        \mathrm{SNR} = \frac{\bm{h}^{*}\bm{s}\bm{s}^{*}\bm{h}}{\bm{h}^{*}\bm{R_{v}}\bm{h}}.
    \end{equation}
    As discussed in \S2.1, \cite{North_1963} showed that a linear matched filter of form: 
    \begin{equation}
        \bm{h} = [\frac{1}{\sqrt{\bm{s}^{*}\bm{R_{v}}^{-1}\bm{s}}}]\bm{R_{v}}^{-1}\bm{s} = C\bm{R_{v}}^{-1}\bm{s},
    \end{equation}
    maximizes the output SNR. The natural question is then how much is the SNR boosted by applying such a filter? This can be analytically calculated for a given signal by comparing the SNR after applying a matched filter with the SNR from applying a flat filter $\bm{1}$ (e.g. a unity matrix of all ones). We start by calculating the SNR after applying the matched filter:
    \begin{equation}
        \mathrm{SNR_{mf}} = \frac{\bm{h}^{*}\bm{s}\bm{s}^{*}\bm{h}}{\bm{h}^{*}\bm{R_{v}}\bm{h}}.
    \end{equation}
    We then substitute for $\bm{h}$, noting that that $\bm{R_{v}}^{-1}$ is Hermitian and therefore $\bm{R_{v}}^{{-1}^{*}} = \bm{R_{v}}^{-1}$:
    \begin{equation}
        \mathrm{SNR_{mf}} = \frac{\bm{R_{v}}^{-1}\bm{s}^{*}\bm{s}\bm{s}^{*}\bm{R_{v}}^{-1}\bm{s}}{\bm{R_{v}}^{-1}\bm{s}^{*}\bm{R_{v}}\bm{R_{v}}^{-1}\bm{s}}.
    \end{equation}
    Under the assumption of uncorrelated noise (which is reasonable for the case of independent interferometric visibilities), there are no off-diagonal terms in $\bm{R_{v}}$ and we can reduce this equation to:
     \begin{equation}
        \mathrm{SNR_{mf}} = \sum_{i}^{N}\|s_i\|^{2}R_{ii}^{-1},
    \end{equation}
    where there are N elements of the signal $\bm{s}$ (which can be summed in multiple dimensions or flattened as shown here). Similarly, if we write the SNR of the flat filter as:
    \begin{equation}
        \mathrm{SNR_{flat}} = \frac{\bm{1}^{*}\bm{s}\bm{s}^{*}\bm{1}}{\bm{1}^{*}\bm{R_{v}}\bm{1}},
    \end{equation}
    then we find it reduces to:
    \begin{equation}
        \mathrm{SNR_{flat}} = \frac{(\sum_{i}^{N} \|s_i\|)^{2}}{\mathrm{tr}[\bm{R_{v}}]}.
    \end{equation}
    So the ratio of these two SNRs, or the total SNR boost from a matched filter, is:
    \begin{equation}
        \mathrm{boost} = \frac{\mathrm{SNR_{mf}}}{\mathrm{SNR_{flat}}} = \frac{(\sum_{i}^{N}\|s_i\|^{2}R_{ii}^{-1})\mathrm{tr}[\bm{R_{v}}]}{(\sum_{i}^{N} \|s_i\|)^{2}}.
    \end{equation}

\section{Calculating SNR boost in comparison to image-plane measurements}
\label{appendix:B}
    The boost value in equation A8 can be analytically calculated for a given filter kernel, but is not particularly useful at this point as it has not been related to the image-plane SNRs discussed throughout the paper. Thus the fundamental problem is how to relate the visibilities to an image-plane SNR. We start by writing down the definition of SNR in the dirty image $\bm{I^{D}}$, or the raw discrete Fourier transform of the visibilites (i.e. not deconvolved):
    \begin{equation}
    	\mathrm{SNR} = \frac{\bm{I^{D}}}{\bm{\Delta I^{D}}} = \frac{\displaystyle\sum_{k} W_{k} V_{k}}{(\displaystyle\sum_{k} W_{k}^2 \sigma_{k}^2)^{1/2}},	    
    \end{equation}
    where there are $k$ visibilities in the dataset, each with a source visibility contribution $V_{k}$, total weight $W_{k}$ (including any taper weights, density weights, and the variance weights $w_k$ discussed in the main text), and thermal noise $\sigma_k$. Notation is borrowed from \cite{Briggs_1995}, which contains a detailed discussion of image-plane SNRs and their relation to the measured visibilities. If the data is gridded into cells $\{p,q\}$ and the noise properties of all visibilities within each cell are similar, an approximate form can be written:
    \begin{equation}
    	\mathrm{SNR} = \frac{\displaystyle\sum_{p,q} W_{pq} V_{pq}}{(\displaystyle\sum_{p,q} W_{pq}^2 \sigma_{pq}^2)^{1/2}}.	    
    \end{equation}
    In particular, we are interested in the SNR at a particular location in the dirty map, e.g. the peak pixel in a given channel. If this pixel is phase shifted to the map center, the SNR can be written as:
    \begin{equation}
        \mathrm{SNR}'(0,0) = \frac{\displaystyle\sum_{p,q} W_{pq} |V_{pq}|}{(\displaystyle\sum_{p,q} W_{pq}^2 \sigma_{pq}^2)^{1/2}}.	    
    \end{equation}
    If we consider a moment-0 map of a resolved source, however, the SNR at the center of the moment map is:
    \begin{equation}
        \mathrm{SNR_{mom0}} = \frac{\displaystyle\sum_{c,p,q} W_{cpq} \mathrm{Re}(V_{cpq})}{(\displaystyle\sum_{c,p,q} W_{cpq}^2 \sigma_{cpq}^2)^{1/2}},	    
    \end{equation}
    and only the projected real component of each visibility will contribute signal. The SNR will then decrease as the ratio of the emission size to the resolution element increases, as discussed in \cite{Crane_1986} and \cite{Yen_2016}. Compounding this issue, if the source has a strong spatio-kinematic signature and peak emission moves throughout the dirty map, then the projected real component will vary as a function of channel $c$. Applying these properties to Equation A8, we can estimate the SNR boost of the matched filter over a peak moment-0 value as:
    \begin{equation}
        \mathrm{boost} = \frac{\mathrm{SNR_{mf}}}{\mathrm{SNR_{mom0}}} = \frac{(\sum_{i}^{N}\|V_i\|^{2}R_{ii}^{-1})\mathrm{tr}[\bm{R_{v}}]}{(\sum_{i}^{N} \mathrm{Re}(V_i))^{2}}.
    \end{equation}    
    Aligning the signal in the image plane through pixel shifting and stacking (e.g. as in \cite{Matra_2015} or \cite{Yen_2016}) is analogous to phase shifting the individual visibilities to the map center. Re($V_{pq}$) can then be replaced by $|V_{pq}|$, and the SNR after applying a pixel shifting method is roughly:
    \begin{equation}
        \mathrm{SNR_{ps}} = \mathrm{SNR'_{mom0}} = \frac{\displaystyle\sum_{c,p,q} W_{cpq} |V_{cpq}|}{(\displaystyle\sum_{c,p,q} W_{cpq}^2 \sigma_{cpq}^2)^{1/2}}.	    
    \end{equation}
    Returning to equation A8 and applying this logic, we can write the boost as:
    \begin{equation}
        \mathrm{boost} = \frac{\mathrm{SNR_{mf}}}{\mathrm{SNR_{ps}}} = \frac{(\sum_{i}^{N}\|V_i\|^{2}R_{ii}^{-1})\mathrm{tr}[\bm{R_{v}}]}{(\sum_{i}^{N} |V_i|)^{2}},
    \end{equation}   
    which defines the additional benefit matched filtering provides over a pixel stacking approach.
    
    Equations B4 and B6 can be applied to any visibility sampled filter kernel to calculate these boosts. We note, however, that due to the line-broadening of most astronomical signals, true phase alignment from a pixel shifting approach is not possible and therefore the boost formulas are only approximations.

\section{Data weights and noise covariance matrices}    
\label{appendix:C}

    \subsection{Interferometric data weights}
        In general, each visibility $V_{i}$ in an interferometric data set corresponding to an antenna pair $(m,n)$ will have a characteristic variance $\sigma_{mn}^{2}$, which is often assumed to be dominated by the system noise \citep[see e.g. Chap. 6 of][]{Thompson_2017}. When the system noise dominates, $\sigma_{mn}$ can be written in units of Jy as:
        \begin{equation}
            \sigma_{mn}(Jy) = \frac{2k}{n_{q} n_{c} A_{eff}} \sqrt{\frac{T_{sys,m} T_{sys,n}}{2\Delta \nu \Delta t}} \times 10^{26},
        \end{equation}
        where $k$ is Boltzmann's constant, $n_q$ and $n_c$ are the quantization and correlator efficiencies, $A_{eff}$ is the effective antenna area, $T_{sys,m}$ and $T_{sys,n}$ are system temperatures for antennas $m$ and $n$, respectively, $\Delta \nu$ is the effective channel bandwidth, and $\Delta t$ is the integration time. For identical antennas and system temperatures, this can be simplified as:
        \begin{equation}
            \sigma_{mn}(Jy) = \frac{2k T_{sys}}{n_{q} n_{c} A_{eff}} \sqrt{\frac{1}{2\Delta \nu \Delta t}} \times 10^{26}.
        \end{equation}
        The data weight for each visibility is then calculated as $w_i=1/\sigma_i^2$. For instruments which record channelized system temperatures (e.g. ALMA), the weights will also be channelized, and are recorded in CASA as a `weight spectrum' for each visibility.\footnote{For more details on how weights are handled in different versions of CASA, see \url{https://casa.nrao.edu/casadocs/casa-5.1.0/reference-material/data-weights}.}
    
    \subsection{Noise covariance matrices}
        As discussed in Section 2.1, a matched filter can be written as:
        \begin{equation}
            \bm{h} = [\frac{1}{\sqrt{\bm{s}^{*}\bm{R_{v}}^{-1}\bm{s}}}]\bm{R_{v}}^{-1}\bm{s} = C\bm{R_{v}}^{-1}\bm{s},
        \end{equation}
        where $\bm{R_{v}}$ is the noise covariance matrix. When the $n_c$ channels in an interferometric dataset are fully independent, $\bm{R_{v}}$ can be written for an individual visibility $V_{i}$ as a $n_c \times n_c$ diagonal matrix:
        \begin{equation}
            \bm{R_{v}} = 
            \begin{bmatrix}
                \sigma_{1}^{2}    &                   &                   &                       \\
                                    & \sigma_{2}^{2}  &                   &                       \\
                                    &                   & \ddots            &                       \\
                                    &                   &                   & \sigma_{n_{c}}^{2}  \\
            \end{bmatrix}
        \end{equation}
        and $\bm{R_v^{-1}}$ is then:
        \begin{equation}
        \bm{R_v^{-1}} = 
        \begin{bmatrix}
            \frac{1}{\sigma_{1}^{2}}    &                   &                   &                       \\
                                & \frac{1}{\sigma_{2}^{2}}  &                   &                       \\
                                &                   & \ddots            &                       \\
                                &                   &                   & \frac{1}{\sigma_{n_{c}}^{2}}  \\
        \end{bmatrix}
        \end{equation}
        i.e. a diagonal matrix initialized to the channelized data weights. When the weights are not channelized, or can be approximated as equal across channels ($w_j \approx w_i~\forall~j$), $R_v^{-1}$ can be written as:
        \begin{equation}
            \bm{R_v^{-1}} = w_{i}\bm{I_{n_c}},
        \end{equation}
        where $\bm{I_{n_c}}$ is an $n_c \times n_c$ identity matrix.
        
        \subsubsection{Correlated channels}
            In the case of fully independent channels, the filter and normalization prefactor are simple to calculate due to the lack of off-diagonal elements in the noise covariance matrix. In practice, however, channels are often correlated. In particular, astronomical interferometric datasets are generally correlated due to the use of a window function (e.g. Hann, Hamming, etc.) to reduce the ringing effect introduced by the finite maximum lag time of the correlator hardware.\footnote{A full description of these effects and the choice of window function for ALMA can be found at \url{https://safe.nrao.edu/wiki/pub/Main/ALMAWindowFunctions/Note_on_Spectral_Response_V2.pdf}.} As the Hann function is a popular choice of window function (and applied directly in the time domain for the ALMA data described in this paper), we derive here the appropriate noise covariances matrices for Hanning smoothed data. Similar results could be calculated for other choices of window function.
            
            In the frequency/channel domain, the Hann window applied to an observation $\bm{x}$ can be written as:
            \begin{equation}
                x'_i = \frac{1}{4}x_{i-1} + \frac{1}{2}x_{i} + \frac{1}{4}x_{i+1}.
            \end{equation}
            For an observation which can be linearly decomposed into signal and additive white gaussian noise ($\bm{x} = \bm{s} + \bm{v}$), we can calculate the noise covariance matrix of the smoothed data, $\bm{R'_{v}}$, which now contains off-diagonal elements:
            \begin{equation}
                R'_{v}[j, k] = E[v'_j v'^{*}_k],
            \end{equation}
            where 
            \begin{equation}
                v'_j = \frac{1}{4}v_{j-1} + \frac{1}{2}v_{j} + \frac{1}{4}v_{j+1}.
            \end{equation}
            Noting that in the uncorrelated case
            \begin{equation}
                E[v_j v^{*}_k] =
                \begin{cases}
                \sigma_{j}^2, & \text{if } j = k \\
                0, & \text{otherwise}
                \end{cases}
            \end{equation}
            we find that for diagonal elements of $R'_{v}$, Eq. C23 reduces to
            \begin{equation}
                R'_{v}[j, j] = \frac{\sigma_{j-1}^2}{16} + \frac{\sigma_{j}^2}{4} + \frac{\sigma_{j+1}^2}{16},
            \end{equation}
            and for the populated off-diagonal elements it reduces to
            \begin{equation}
                R'_{v}[j, j\pm1] = \frac{\sigma_{j}^2}{8} + \frac{\sigma_{j\pm1}^2}{8},
            \end{equation}
            and
            \begin{equation}
                R'_{v}[j, j\pm2] = \frac{\sigma_{j\pm1}^2}{16}.,
            \end{equation}
            Under the assumption that the channelized weights are all approximately equal for a given visibility ($w_j \approx w_i~\forall~j$), $\bm{R'_{v}}$ can be written as
            \begin{equation}
            \bm{R'_{v}} \approx \sigma_{i}^{2}
            \begin{bmatrix}
                \frac{3}{8}     & \frac{1}{4}   & \frac{1}{16}  &               &              \\
                \frac{1}{4}     & \frac{3}{8}   & \ddots        & \ddots        &              \\
                \frac{1}{16}    & \ddots        & \ddots        & \ddots        & \frac{1}{16} \\
                                & \ddots        & \ddots        & \frac{3}{8}   & \frac{1}{4}  \\
                                &               & \frac{1}{16}  & \frac{1}{4}   & \frac{3}{8}  \\
            \end{bmatrix}
            = \sigma {'_{i}}^{2}
            \begin{bmatrix}
                1           & \frac{2}{3}   & \frac{1}{6}   &               &              \\
                \frac{2}{3} & 1             & \ddots        & \ddots        &              \\
                \frac{1}{6} & \ddots        & \ddots        & \ddots        & \frac{1}{6} \\
                            & \ddots        & \ddots        & 1             & \frac{2}{3}  \\
                            &               & \frac{1}{6}   & \frac{2}{3}   & 1             \\
            \end{bmatrix}
            = \sigma {'_{i}}^{2} ~ [\bm{M}]
            \end{equation}
            where $\sigma {'_{i}} = \sqrt{3/8} \sigma_{i}$. As tasks such as \texttt{statwt} in CASA do not consider effective channel bandwidth, ${w'}_{i} = 1 / \sigma {'_{i}}^2$ is likely what will be reported as the data weights of the observations.\footnote{See \url{https://casaguides.nrao.edu/index.php/DataWeightsAndCombination} for details about the absolute accuracy of data weights in CASA.} The assumption of equal weights across channels is reasonable when $T_{sys}$ is stable across channels, which will likely be true unless there were issues with data calibration or there are strong water lines in the data. Under this assumption, the matrix inversion only needs to be computed once:
            \begin{equation}
            \bm{{R'_{v}}^{-1}} \approx \frac{1}{\sigma {'_{i}}^{2}} ~ [\bm{M}]^{-1}
            = {w'}_{i} ~ [\bm{M}]^{-1}.
            \end{equation}
            When the data is initially Hanning smoothed, the edge channels are clipped. Thus for the purposes of preventing boundary effects during inversion, the covariance matrix can be treated as describing a infinitely wide dataset. In practice, we extend the matrix to be several times larger than $n_c$, and then window a $n_k$ sized portion from the center of $\bm{{R'_{v}}^{-1}}$. In the unbinned case, however, $\bm{R'_{v}}$ is an ill-conditioned matrix (the condition number is ~$>$~$10^{10}$ for a typical ALMA spectral window), making the matrix inversion numerically unstable. The issue can be avoided by channel binning the data by a factor of two, as discussed in the following section.
            
        \subsubsection{Averaged correlated channels}
            As Hanning smoothing inflates effective channel width by a factor of 8/3, it is very common to bin data across channels by a factor of 2; this is the default for ALMA observations. Here we calculate the appropriate covariance matrices for binning factors of 2, 3, and 4. A similar method can be followed for other binning factors, but we note that past a binning factor of 4, the covariance matrix and its inverse rapidly approach the uncorrelated case and the channel correlation can likely be neglected without significant adverse effect.
            
            With a binning factor of 2 applied to data that has been Hanning smoothed, the data (indexed by channel in the averaged dataset) can be described as:
            \begin{equation}
                x'_{i,bin\times2} = \frac{(\frac{1}{4}x_{i-0.5} + \frac{1}{2}x_{i} + \frac{1}{4}x_{i+0.5}) + (\frac{1}{4}x_{i} + \frac{1}{2}x_{i+0.5} + \frac{1}{4}x_{i+1})}{2} = \frac{1}{8}x_{i-0.5} + \frac{3}{8}x_{i} + \frac{3}{8}x_{i+0.5} + \frac{1}{8}x_{i+1},
            \end{equation}
            and the noise component is therefore:
            \begin{equation}
                v'_{i,bin\times2} = \frac{1}{8}v_{i-0.5} + \frac{3}{8}v_{i} + \frac{3}{8}v_{i+0.5} + \frac{1}{8}v_{i+1}.
            \end{equation}
            Substituting this into Eq. C23 and applying Eq. C25, we find that for diagonal elements of $\bm{R'_{v}}$
            \begin{equation}
                R'_{v,bin\times2}[j, j] = \frac{\sigma_{j-0.5}^2}{64} + \frac{9\sigma_{j}^2}{64} + \frac{9\sigma_{j+0.5}^2}{64} + \frac{\sigma_{j+1}^2}{64},
            \end{equation}
            and the only populated off-diagonal elements are
            \begin{equation}
                R'_{v,bin\times2}[j, j\pm1] = \frac{3\sigma_{j\pm0.5}^2}{64} + \frac{3\sigma_{j\pm1}^2}{64}.
            \end{equation}
            Assuming the weights are roughly equivalent across channels for a given visibility, $\bm{R'_{v}}$ can be approximated as
            \begin{equation}
            \bm{R'_{v,bin\times2}} \approx \sigma_{i}^{2}
            \begin{bmatrix}
                \frac{5}{16}     & \frac{3}{32}   &               &              \\
                \frac{3}{32}     & \frac{5}{16}   & \ddots        &              \\
                                & \ddots        & \ddots        & \frac{3}{32}  \\
                                &               & \frac{3}{32}   & \frac{5}{16}  \\
            \end{bmatrix}
            = \sigma {'_{i}}^{2}
            \begin{bmatrix}
                1     & \frac{3}{10}   &               &              \\
                \frac{3}{10}     & 1   & \ddots        &              \\
                                & \ddots        & \ddots        & \frac{3}{10}  \\
                                &               & \frac{3}{10}   & 1  \\
            \end{bmatrix}
            = \sigma {'_{i}}^{2} ~ [\bm{M_{bin\times2}}]
            \end{equation}
            where $\sigma {'_{i}} = \sqrt{5/16} \sigma_{i}$. Correspondingly,
            \begin{equation}
            \bm{{R'_{v,bin\times2}}^{-1}} \approx \frac{1}{\sigma {'_{i}}^{2}} ~ [\bm{M_{bin\times2}}]^{-1}
            = {w'}_{i} ~ [\bm{M_{bin\times2}}]^{-1}.
            \end{equation}
            where the weights ${w'}_{i}$ were calculated from the binned data using a task such as \texttt{statwt}. In contrast to the unbinned case, $\bm{R'_{v}}$ is now tridiagonal and well-conditioned for inversion.
            
            Repeating these calculations for binning factors of 3 and 4, we find that:
            \begin{equation}
            \bm{R'_{v,bin\times3}} \approx \sigma_{i}^{2}
            \begin{bmatrix}
                \frac{1}{4}     & \frac{1}{24}   &               &              \\
                \frac{1}{24}     & \frac{1}{4}   & \ddots        &              \\
                                & \ddots        & \ddots        & \frac{1}{24}  \\
                                &               & \frac{1}{24}   & \frac{1}{4}  \\
            \end{bmatrix}
            = \sigma {'_{i}}^{2}
            \begin{bmatrix}
                1     & \frac{1}{6}   &               &              \\
                \frac{1}{6}     & 1   & \ddots        &              \\
                                & \ddots        & \ddots        & \frac{1}{6}  \\
                                &               & \frac{1}{6}   & 1  \\
            \end{bmatrix}
            = \sigma {'_{i}}^{2} ~ [\bm{M_{bin\times3}}]
            \end{equation}
            where $\sigma {'_{i}} = \sqrt{1/4} \sigma_{i}$,
            \begin{equation}
            \bm{{R'_{v,bin\times3}}^{-1}} \approx \frac{1}{\sigma {'_{i}}^{2}} ~ [\bm{M_{bin\times3}}]^{-1}
            = {w'}_{i} ~ [\bm{M_{bin\times3}}]^{-1},
            \end{equation}
            and
            \begin{equation}
            \bm{R'_{v,bin\times4}} \approx \sigma_{i}^{2}
            \begin{bmatrix}
                \frac{13}{64}     & \frac{3}{128}   &               &              \\
                \frac{3}{128}     & \frac{13}{64}   & \ddots        &              \\
                                & \ddots        & \ddots        & \frac{3}{128}  \\
                                &               & \frac{3}{128}   & \frac{13}{64}  \\
            \end{bmatrix}
            = \sigma {'_{i}}^{2}
            \begin{bmatrix}
                1     & \frac{3}{26}   &               &              \\
                \frac{3}{26}     & 1   & \ddots        &              \\
                                & \ddots        & \ddots        & \frac{3}{26}  \\
                                &               & \frac{3}{26}   & 1  \\
            \end{bmatrix}
            = \sigma {'_{i}}^{2} ~ [\bm{M_{bin\times4}}]
            \end{equation}
            where $\sigma {'_{i}} = \sqrt{13/64} \sigma_{i}$,
            \begin{equation}
            \bm{{R'_{v,bin\times4}}^{-1}} \approx \frac{1}{\sigma {'_{i}}^{2}} ~ [\bm{M_{bin\times4}}]^{-1}
            = {w'}_{i} ~ [\bm{M_{bin\times4}}]^{-1}.
            \end{equation}

\section{Filter kernel generation}
\label{appendix:D}
    The Keplerian filter kernel presented in Fig. \ref{Figure 2} was calculated for the viewing geometry of the protoplanetary disk around T Tauri star TW Hya \citep[with an inclination of 7$\degree$ and PA of 155$\degree$; e.g.,][]{Qi_2004, Andrews_2012}. The Keplerian velocity field is calculated as
    \begin{equation}
        v = \sqrt{\frac{G M_*}{r}},
    \end{equation}
    with an assumed stellar mass of 0.8 M$_{\odot}$ \citep{Hughes_2011}. Using this field, we computed the emitting region of the disk for channels with 0.2~km~s$^{-1}$ spacing.
    
    The parametric model filter kernel was calculated from the `fiducial' model in \cite{Walsh_2016}, where CH$_3$OH around TW Hya is constrained to a vertical layer z/r $\textless$ 0.1 between radii of 30-100 AU. From this abundance structure, an emission profile was calculated for the 3$_{12}$-3$_{03}$ transition using LIME. As seen in Fig. \ref{Figure 2}, the emission from this model tapers radially due to decreasing column density and temperature, in contrast to the Keplerian mask which simply has a hard outer radius cutoff. Additionally, the CH$_3$OH model has an inner disk depletion (as CH$_3$OH is mainly formed through hydrogenation of CO on grain surfaces outside the CO snowline), which is not present in the simple Keplerian model.
    
    Finally, the data-driven filter kernel was generated from observations of H$_2$CO around TW Hya \citep{Oberg_2017}. The data were imaged in CASA using \texttt{CLEAN} with natural weighting, yielding a high SNR image cube. After imaging, all noise below 3$\sigma$ and any emission outside of a 3$\arcsec$ radius were masked out, creating a mostly noiseless approximation of the true H$_2$CO emission distribution. $(u,v)$ plane kernels were generated from each of these image cubes using \texttt{vis$\textunderscore$sample}, as described in \S2.2, \S2.3, and \S2.4.
    
\end{document}